\documentclass[twocolumn,eqsecnum,superscriptaddress,showpacs]{revtex4}
\usepackage{amsmath,amssymb}
\usepackage{bm}

\begin{document}
\title{Problems of Real Scalar Klein-Gordon Field}% Force line breaks with \\

\author{Sachiko \textsc{Oshima}} 
\affiliation{Department of Physics, Faculty of Science, Tokyo Institute 
of Technology, Tokyo, Japan} 
\author{Seiji \textsc{Kanemaki}} \email{kanemaki@phys.cst.nihon-u.ac.jp}
\author{Takehisa \textsc{Fujita}} \email{fffujita@phys.cst.nihon-u.ac.jp}
\affiliation{
Department of Physics, Faculty of Science and Technology
Nihon University, Tokyo, Japan 
}%

\date{\today}% It is always \today, today, but you may specify any date with \date.

\begin{abstract}
We examine the negative energy solution in Klein-Gordon equation  
in terms of the number of field components.  A scalar field 
has only one component, and there is no freedom left for an anti-particle since 
the Klein-Gordon equation failed to take the negative energy solution into account. 
This is in  contrast to the Dirac equation which has four components of fields. 
It is shown that the current density for a real scalar field is always zero 
if the field is classical, but infinite if the field is quantized. 
This suggests that the condition of a real field  must be physically too strong. 

\end{abstract}


\maketitle

\section{ Introduction}
When one treats a particle which moves with its velocity 
close to the velocity of light, then one should employ the relativistic 
kinematics. The Einstein relation for a particle with its mass $m$ is given as
$$ E=\sqrt{m^2+{\bf p}^2 } . \eqno{(1.1)} $$
For the quantized equations of this particle, there are two equations,  
that is, Klein-Gordon and Dirac equations. 

The equation for relativistic bosons can be described by the Klein-Gordon equation, 
$$ \left({\partial^2\over{\partial t^2}}-\nabla^2 +m^2 \right) \phi (x)= 0  
\eqno{(1.2)}  $$
where we denote $x=(t, {\bf r})$. 

It should be important to note that the boson field $\phi$ has only one component, 
and therefore it is believed to describe the spin zero particle. In this case, however, 
it is interesting to ask where the freedom of the negative energy state may appear. 
Since the boson field $\phi$ has only one component, it is difficult to 
imagine that the effects of the negative energy solutions are properly 
taken into account. This means that we should find out where the freedom 
of the anti-particle should appear. In this respect, the boson field $\phi$ is 
essentially the same as the Schr\"odinger field $\psi$ in the non-relativistic 
quantum mechanics. At least, classically, the boson field $\phi$ which should be 
complex like the Schr\"odinger field $\psi$ should correspond to one boson state. 

Therefore, if we could find an equation which has two components of field, then 
we should be able to understand where the degree of freedoms concerning the negative 
energy state appears. However, as long as we carry out the procedure of factoring 
out eq.(1.2) into equations with two component fields, we see that it is impossible. 
As is well known, the equations with four component fields are only possible, which 
is the Dirac equation \cite{q1}. 

There is an interesting attempt to construct field equations of bosons with two 
component spinor by Gross \cite{gross}. However, the Hamiltonian becomes non-Hermite even 
though the right dispersion relation for bosons is obtained. 

On the other hand, 
there is one example that has the proper two components for boson fields. 
That is, the gauge field in QED. The vector field $\bm{A}$ has the two components 
and obeys coupled equations which have some similarities with the Klein-Gordon 
equation with some constraints.  This indicates that there a vector field 
with a finite mass may have some difficulty since, in this case, the degree of freedom 
cannot be properly taken into account in the coupled equations as the massive 
elementary vector boson. 

Here, several questions may arise. The first question is,  what is the anti-particle of 
a real scalar field ?   Is it related to the negative energy solutions ? 
The second question is connected to the current density of a real scalar field. 
Since it is easy to prove that the current density of a real scalar field vanishes,  
what is the physics of no current density of a scalar field ? 
As the third question, what is the physical meaning of a real scalar field ? 
In the  Schr\"odinger field $\psi$, it is naturally complex. If the field $\psi$ 
is real, it does not depend on time and the energy eigenvalue must be zero, and 
it becomes unphysical. Therefore, if a real scalar field exists in Klein-Gordon equation, 
then how can one make a non-relativistic limit of the Klein-Gordon real scalar field ? 

Here, we discuss the above questions and examine whether a real scalar 
field can exist as a physical observable or not in Klein-Gordon equation. 
In most of the field theory textbooks, we find that pion with the positive charge 
is an anti-particle of pion with the negative charge. This can be understood easily 
if we look into the structure of the pion in terms of quarks. $\pi^{\pm}$ are indeed 
anti-particle to each other by changing quarks into anti-quarks. 

Since pions are not an elementary particle, their dynamics must be 
governed by the complicated quark dynamics. Under some drastic approximation, 
the motion of pion may be governed by the Klein-Gordon equation if one is only 
interested in the center of motion of pion. 

\section{Negative energy state}

It looks that eq.(1.2) contains the negative energy state. 
However, one sees that eq.(1.2) is only one component equation and, therefore 
the eigenvalue of $E^2$  can be obtained as a physical observable. 
There is no information from the Klein-Gordon equation for the energy $E$ itself, but 
only $E^2$ as we see it below, 
$$  \left(  -{\mbox{\boldmath $\nabla$}}^2+m^2\right) 
\phi = E^2 \phi . \eqno{(2.1)}  $$
Thus, one obtains only one information from the Klein-Gordon equation. 
In this respect, there should be no negative energy physical state 
in the Klein-Gordon equation. 

After one obtains the value of $E^2$, one may say that one finds two values 
for $E$. However, the way one obtains two solutions has nothing to do with 
physics.  Physically, the positive solution of eq.(2.1) 
should be taken, and this is just like the determination of a radius of a circle 
in elementary analysis. When the radius $r$ is given as $r^2 =1$, for example, 
then one should take  $r=1$ solution. 

Mathematically, the wave function may include the solutions with 
$E= \pm \sqrt{m^2+{\bm{k}}^2 } $. 

\subsection{Classical field}
Eq. (2.1) has two independent solutions and can be given by requiring 
that the state should be the eigenstate of momentum
$$ {\phi} (x) = C_+e^{-i\omega_k t+i{\bf k}\cdot{\bf r} }, \ \ \ \ 
C_- e^{i\omega_k t-i{\bf k}\cdot{\bf r} }  \eqno{(2.2)} $$
where $\omega_k=\sqrt{k^2+m^2} $. This shape of the solution is determined 
since the momentum $p^\mu=(\omega, \bm{k})$ and the coordinate 
$x^\mu=(t, \bm{r})$ can make a Lorentz scalar as 
$$ p^\mu x_\mu =\omega_k t-{\bf k}\cdot{\bf r}  $$
which is just the one found in eq.(2.2). 

Now, we discuss the current density 
of the Klein-Gordon field which is defined as 
$$ \rho(x) ={i}\left(\phi^\dagger(x) {\partial \phi (x)\over{\partial t}}-
{\partial \phi^\dagger(x)\over{\partial t}}\phi(x) \right)  
\eqno{(2.3a)} $$
$$ \bm{j}(x) =-{i}\left[\phi^\dagger(x) (\nabla \phi (x))-
(\nabla \phi^\dagger(x))\phi(x) \right] . 
\eqno{(2.3b)} $$
It should be noted that the current density must be hermitian and therefore 
the shape of eqs.(2.3) is uniquely determined. One cannot change the order 
between 
$$ {\partial \phi^\dagger (x)\over{\partial t}} \ \ \  {\rm and} \ \ \ \phi(x)  $$ 
in the second term of eq.(2.3a). 

Now, we come to an important observation that a real scalar field should have 
a serious problem. The real scalar field $\phi (x)$ can be written as 
$$ {\phi} (x) =\sum_{\bf k} {1\over{\sqrt{2V\omega_k}}}\left[ 
a_ke^{-i\omega_k t+i{\bf k}\cdot{\bf r} }+
a_k^\dagger e^{i\omega_k t-i{\bf k}\cdot{\bf r} } \right] \eqno{(2.4)} $$
where $V$ denotes the box. We assume that the $\phi (x)$ is still a classical 
field, that is, $a_k^\dagger $ and $a_k$ are not operators, but just the same 
as $C_{\pm}$ in eq.(2.2). 

In this case, it is easy to prove that the current densities of $\rho (x)$ and 
$ \bm{j}(x)$ which are constructed from the real scalar field $\phi (x)$ vanish to zero
$$  \rho(x) = 0, \ \ \ \  \bm{j}(x)=0 . \eqno{(2.5)} $$
This means that there is no flow of the real scalar field, at least, classically. 
This is clear since a real wave function in the Schr\"odinger equation cannot propagate. 
Therefore, the condition that the scalar field should be a real field must be 
physically too strong. Even in the Schr\"odinger field, one cannot require that 
the field should be real as will be discussed in section IV.  

In other words, the scalar field $\phi$ can be a real if it is a solution 
that satisfies the boundary conditions, but one cannot require that it 
should be a real field. 

\subsection{Quantized field}
Now, we come to  the current density when the field is quantized.  
Below it is shown that the current density of the real scalar field has some problem 
even if quantized, contrary to a common belief.  

When we quantize the boson field of $\phi$, then  $a_k^\dagger $ and $a_k$ become 
creation and annihilation operators 
$$ \hat{\phi} (x) =\sum_{\bf k} {1\over{\sqrt{2V\omega_k}}}\left[ 
a_ke^{-i\omega_k t+i{\bf k}\cdot{\bf r} }+
a_k^\dagger e^{i\omega_k t-i{\bf k}\cdot{\bf r} } \right]. \eqno{(2.6)} $$
In this case, the current density of eq.(2.4) becomes
$$ \hat{\rho} ={i}\left(\hat{\phi}(x) \hat{\Pi}(x)-
\hat{\Pi}(x)\hat{\phi}(x) \right) = 
{i}[\hat{\phi}(x), \hat{\Pi}(x) ]  \eqno{(2.7)} $$
where $\hat{\Pi}(x)$ is a conjugate field of $\hat{\phi}(x)$, that is, 
$\Pi(x)={\dot \phi}(x)$. 
However,  the quantization condition of the boson fields with eq.(2.6) becomes 
$$ [\hat{\phi}({\bf r}), \hat{\Pi}({\bf r}') ]_{t=t'} =i \delta ({\bf r}-{\bf r}') . 
\eqno{(2.8)} $$
Therefore, eq.(2.7) becomes
$$ \hat{\rho} ={i}[\hat{\phi}(x), \hat{\Pi}(x) ] 
=-  \delta ({\bf 0}) . \eqno{(2.9)} $$
Thus, the current density of the quantized real boson field becomes infinity 
after the quantization ! In this sense, the current density of a real scalar field 
behaves completely in an unphysical manner. 
Therefore, it is by now obvious that the current density of the real scalar field 
has  an improper physical meaning. This should be related to the fact 
that  a real scalar field condition  is too strong and one cannot impose 
this real condition on  $\phi$  at the Lagrangian density level. 

\section{Complex scalar fields}
Since a real scalar field has a difficulty not only in the classical case but 
also in the quantized case, it should be worth considering a complex scalar field. 
In this case, the complex scalar field can be written as
$$ {\phi} (x) =\sum_{\bf k} {1\over{\sqrt{2V\omega_k}}}\left[ 
a_ke^{-i\omega_k t+i{\bf k}\cdot{\bf r} }+
b_k^\dagger e^{i\omega_k t-i{\bf k}\cdot{\bf r} } \right].  \eqno{(3.1)} $$
In this case, the current density is well defined and has no singularity 
when quantized. 

According to a common belief, the complex scalar field should describe 
charged bosons, one which has a positive charge and the other which has a negative 
charge. But a question may arise as to where this degree of freedom comes from ? 
By now, we realize that there is no negative energy solution in the Klein-Gordon 
equation. If one took into account the negative energy solution, then one 
should have had the field equations of two components. On the other hand, 
eq.(3.1) assumes a scalar field with two components, and not the result 
of the field equations.  It is therefore most important to seek for 
the two component Klein-Gordon like equation which should be somewhat similar 
to the Dirac equation. 

In any case, if it is a simply complex scalar field, then it should describe 
one boson state, at least, as a classical field. It is therefore surprising 
that the complex scalar field can describe charged particles when quantized.

\section{ Schr\"odinger field }

It is worthwhile discussing the  Schr\"odinger field $\psi(\bm{r},t)$ 
in the non-relativistic 
quantum mechanics. Apart from the kinematics, the Schr\"odinger field should 
have the same behavior as the scalar field in the classical field theory. 
The Lagrangian density of the 
Schr\"odinger field can be written 
$$ {\cal L} = i \psi^\dagger   {\partial \psi  \over{\partial t}} -
 {1 \over 2m}   \bm{\nabla} \psi^\dagger \bm{\nabla} \psi
  -\psi^\dagger U \psi . \eqno{(4.1)} $$
We note that the Lagrangian density of eq.(4.1) should have the $U(1)$ symmetry. 

From eq.(4.1), we obtain the  Schr\"odinger equation  
$$ i{\partial \psi(\bm{r},t)\over{\partial t}} = 
\left(-{1\over 2m}\nabla^2 + U \right)\psi(\bm{r},t)  . \eqno{(4.2)} $$ 
The solutions of eq.(4.2) in the  $U=0$ case  can be obtained in a box with its volume $V$, 
$$ \psi_1 ={C_1\over{\sqrt{ V}}}e^{-iE_{\bm{k}}t} e^{i\bm{k}\cdot \bm{r} }, \ \ \ 
\psi_2 ={C_2\over{\sqrt{ V}}}e^{-iE_{\bm{k}}t} e^{-i\bm{k}\cdot \bm{r} } \eqno{(4.3)}  $$
where $E_{\bm{k}}={ \bm{k}^2\over 2m}$ and 
$\bm{k}$ denote  quantum numbers which should  correspond to the momentum 
of a particle. 

\subsection{Unphysical real field condition }
It should be important to note that the complex Schr\"odinger field just 
corresponds to one particle state. In fact,  
if one imposes the condition that the field $\psi (\bm{r},t)$ should be a real function, 
that is 
$$ \psi (\bm{r},t) = \psi^\dagger (\bm{r},t) \eqno{(4.4)} $$
then one obtains from the Schr\"odinger equation
$$ {\partial \psi(\bm{r},t)\over{\partial t}} =0 . \eqno{(4.5)} $$
That is, the  Schr\"odinger field becomes time-independent and in addition, 
the energy eigenvalue $E$ is zero
$$ E=0  \eqno{(4.6)} $$
since 
$$ \left(-{1\over 2m}\nabla^2 + U \right)\psi(\bm{r}) =0 . \eqno{(4.6)} $$ 
Therefore, the field cannot propagate. It is by now clear that the real field 
condition of $\psi$ is not physically acceptable. 

Since the Klein-Gordon field should be reduced to the  Schr\"odinger field 
in the non-relativistic limit, the real field condition of the Klein-Gordon 
field must also be unphysical.

\section{Elementary bosons}
For practical problems, $ \pi^{\pm}$ mesons are considered to be anti-particles 
to each other. However, they are not  elementary bosons, and one sees  
that, in terms of the quark and anti-quark terminology, 
they are mutually anti-particles to each other. 

As for elementary bosons, there are  $W^{\pm}$ gauge bosons. However,  
they are not anti-particles to each other. Instead, they are 
different particles with different isospin components and they have nothing 
to do with negative energy solutions of the gauge field equations.  
This is just consistent with the present interpretation of the relativistic bosons. 

Further, one often introduces  complex scalar fields, and one says 
that $\phi$ and $\phi^*$ correspond to anti-particles to each other. 
But this is completely different from the anti-particle of fermions where 
the degree of freedoms for the antiparticle exists as the negative energy 
solution in the Dirac equation. 

Indeed, the Dirac equation is quite different from the Klein -Gordon equation. 
The Dirac equation becomes
$$ \left( -i\nabla \cdot {\mbox{\boldmath $\alpha$}}
 +m \beta \right) \psi =E \psi  . \eqno{(5.1)}  $$
Clearly, the Dirac equation is the eigenvalue equation for the energy $E$, and 
therefore the $E$ itself must be physical observables. This indicates that the negative 
energy states must be physically present, and indeed this is just the way to construct 
the physical vacuum state in fermion field theory models \cite{q1,p5}.

\section{Photon}

There is one example which has the right degree of freedom of boson field. That 
is the electromagnetic field $\bm{A}$. The vector field $\bm{A}$ has only the transverse 
components ( $\nabla \cdot \bm{A} =0$ ) and has indeed two components as a boson 
field. In this respect, the gauge field takes into account the negative energy 
degree of freedom in a proper way, though it is realized as a spin degree of freedom. 

There are massive gauge particles which are weak bosons. The weak gauge fields 
acquire their mass by the Higgs mechanism in which the gauge fixing should be done 
together with the spontaneous symmetry breaking for the Higgs fields. 
The problem of the Higgs mechanism is that the gauge fixing is done 
at the Lagrangian density level, and therefore, after the gauge fixing, 
the Lagrangian density 
has no gauge freedom any more. On the other hand, in the normal circumstance, the gauge 
fixing should be made when one wishes to solve the equation of motion since 
the gauge field has a redundancy and the number of freedom should be reduced. 

\section{Spontaneous symmetry breaking}

In the discussion of the spontaneous symmetry breaking in boson field theory model, 
Goldstone presented a complex scalar field model with a double well potential. 

The Lagrangian density of the complex scalar field $\phi(x)$ in the double well potential 
can be written as 
$$  {\cal L} = {1\over 2}  \partial_\mu\phi^\dagger \partial^\mu\phi
 -u_0 \left(  |{\phi}|^2 -\lambda^2 \right)^2    \eqno{ (7.1)} $$
where $u_0$ and $\lambda$ are constant and the Lagrangian density has a $U(1)$ symmetry. 

Now, one rewrites the complex field as 
$$ \phi(x) = (\lambda + \eta(x)) e^{i{\xi(x)\over{\lambda}}} \eqno{(7.2)} $$
where $\eta$ is assumed to be much smaller than the $\lambda$, 
$$ |\eta(x)| << \lambda . $$
In this case, one can rewrite eq.(7.1) as
$$ {\cal L} = {1\over 2}  \partial_\mu \eta \partial^\mu\eta +
{1\over 2}  \partial_\mu \xi \partial^\mu\xi
+U\left( |\lambda + \eta(x)| \right) +... \eqno{(7.3)} $$
Here, one finds the massless boson $\xi$ which is associated 
with the degeneracy  of the vacuum energy. 

This is the spontaneous symmetry breaking  which is indeed found by Goldstone, and 
he pointed out that there should appear a massless boson associated with the symmetry 
breaking \cite{gold,gold2,nova}. 
The degeneracy of the potential vacuum is converted into a massless 
boson degree of freedom. This looks plausible, and at least approximately 
there is nothing wrong with this treatment of the spontaneous symmetry breaking 
phenomena. However, eq.(7.3) is written in terms of two real scalar fields, and 
physically as we discussed above, the real scalar fields cannot correspond to 
any physical particles. 

In addition, if it is a realistic Lagrangian density for a 
 physical massless boson, then its Lagrangian density must be always written as
$$  {\cal L} = {1\over 2}  \partial_\mu\phi^\dagger \partial^\mu\phi  
\eqno{(7.4)} $$
which is composed of the complex scalar field.

\section{Summary}
In nature, all the elementary particles are composed of fermions, except 
gauge bosons. For this statement, people may claim that the Higgs particles 
must be elementary bosons. However, up to now, these particles are not yet 
discovered in nature, and it should be quite interesting to know 
whether there should exist elementary bosons apart from gauge bosons or not. 

The present study shows that there is no 
anti-particle corresponding to the negative energy solution 
since its degree of freedom does not exist. 
Further, the current density of a real scalar field $j_\mu$ is identically zero 
as a classical field theory and diverges as a quantum field theory. In the 
Schr\"odinger field, a real field condition leads to an unphysical state, and 
since the Klein-Gordon field should be reduced to the Schr\"odinger field 
in the non-relativistic limit, a real scalar field condition in the Klein-Gordon 
field cannot be justified, which is consistent with the vanishing 
current density.    

The present study of the Klein-Gordon field strongly suggests that 
the Klein-Gordon field equation itself should not be physically acceptable. 
The first quantization condition of  $[x_i,p_j]=i\hbar\delta_{ij} $ 
is the basic assumption of deriving the Klein-Gordon equation, and therefore 
if the first quantization is not the fundamental principle, one cannot derive 
the Klein-Gordon equation any more. This point should be reexamined in future. 

\vspace{0.1cm}

We would like to thank K. Fujikawa, M. Hiramoto, T. Nihei and H. Takahashi for  
useful discussions.

\end{document}